# Technologies for tunable gamma-ray lenses

Niels Lund

DTU Space, Technical University of Denmark

Lundtoftevej 327, DK 2800, Kgs. Lyngby, Denmark

nl@space.dtu.dk

**Abstract.** The tunable gamma-ray lens has turned out to be a promising alternative to the classical fixed-energy Laue-lenses discussed in the past. We describe here our development work on a miniature pedestal with one-axis tilt adjustment. We also outline our design for an optical system, capable of monitoring the alignment of the many crystals needed. An added benefit of the tunable crystal pedestal is that it relieves both the demands for high precision in the crystal mounting and the stringent requirements for long-term stability of the support platform on which the crystals are mounted. Moreover, mounting the individual crystals on separate pedestals facilitates the use of double layers of crystals.

## 1) Introduction

'Laue-lenses' are gamma-ray telescopes based on Bragg diffraction in crystals. Laue-lenses can only cover a quite limited energy range (a factor two or three) with reasonable sensitivity. In a previous paper (Lund 2020 - hereafter paper-1) we have demonstrated the large gain in energy coverage, which would result if we can retune a Laue lens for different energy bands. However, there are two obvious problems with the retuning idea: one is that the focal length of the lens changes proportionally to the chosen energy range, and the other is the need to readjust precisely the inclination angle of the many thousand crystals in the lens.

Present day dual satellite formation flying technology offers a solution to the first of these problems: the relevant separations of the gamma-ray lens and the detector - 50 to 300 m – are in a manageable range for satellite orbits around the Sun-Earth Lagrange points, and even for highly elliptical orbits with periods of several days. The station-keeping accuracies needed are moderate – in the few mm-range.

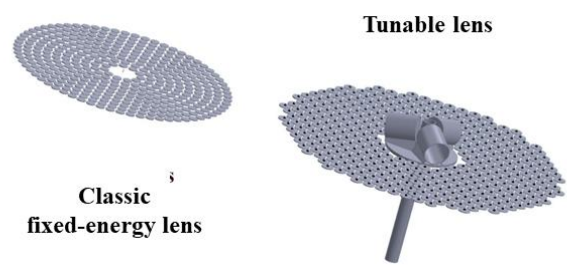

**Fig 1** *Conceptual comparison of the classical fixed-energy Laue lens and the proposed tunable lens. In the case of the tunable lens the precise alignment of the individual Laue-crystals are verified by optical means through an autocollimator system. Three autocollimator telescopes are located on a rotating platform in the center of the tunable lens. It is the rotation axis of this platform, which defines the optical axis of the gamma ray telescope. (Drawings not to scale)*

### 2)   Requirements for the crystal alignment

The second technical problem for the realization of a tunable lens, readjusting and verifying the inclinations of the many thousand crystals is the subject of this paper. In Figure 2A and B we define the two angles which must be kept under control. (The third angle corresponds to rotations around the normal to the Bragg planes, and such rotations do not affect the diffracted beams).

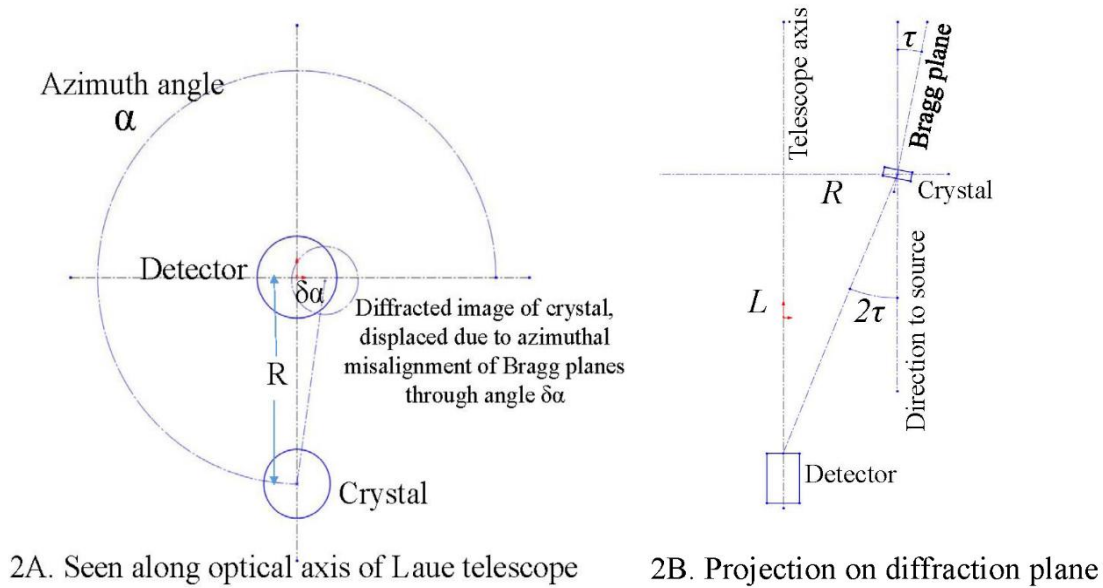

2A. Seen along optical axis of Laue telescope        2B. Projection on diffraction plane

**Fig 2** *Illustration of the two critical alignment angles, α and τ, for the crystals in a Laue lens*

The least critical angle is the azimuthal angle, α, shown in Figure 2A. For every crystal in the lens the normal to the Bragg plane should ideally lie in the plane defined by the telescope axis and the vector from the telescope axis to the center of the crystal. Small deviations, of magnitude $\delta\alpha$, from this plane will displace the diffracted image of the crystal 'sideways' on the detector by an amount, $R \times \delta\alpha$, where $R$ is the distance from the lens optical axis to the crystal. Note that this effect is independent of the focal length of the Laue lens. For the dimensions of the lens considered in this paper $R$ is in the range 0.6 to to 1.8 m. For a detector diameter of 50 mm and a crystal diameter of 40 mm we can tolerate sideways displacements of the diffracted image of a few mm, corresponding to errors on the α-angle of a few arc-minutes. We assume that this accuracy will be achieved by careful design and initial alignment. Consequently, we do not plan for in-flight tuning of the α-angle.

Much more critical is the crystal tilt angle (diffraction angle), τ, illustrated in Figure 2B. Here the purpose of the adjustment is to assure that the diffracted image of the center of all crystals crosses the telescope axis at the same distance, $L$, from the lens. ($L$ is the desired focal length). The diffracted image will fall in the right place if

$$\tau = \arctan(R/2L)$$

The error margin on τ is set by the diameter of the detector relative to $L$. So it appears, that the requirements for the precision of the adjustment will increase linearly with $L$. However, as we shall see, this is not the case.

We usually think about the Bragg planes as something precisely defined within the crystal. But this is not the case for mosaic crystals such as those we propose to use in our lens. The macroscopic mosaic crystal contains an ensemble of close-lying Bragg planes distributed over an angular range. This is precisely what is meant by the 'mosaic width', $\omega$, of the crystal. (We have assumed a value of 30 arc-seconds for $\omega$). The diffraction efficiency at a specific energy, $E$, is at its maximum for a certain direction, but for other, slightly deviant, directions the efficiency drops off at a certain rate – we assume a Gaussian shape of width, $\omega$. So actually, as long as we manage to adjust the crystal tilt relative to the celestial source within an accuracy of a few arc-seconds the crystal will diffract photons at the selected energy at almost full efficiency.

Note that the diffracted image of a crystal at the detector level for the energy, $E$, remains fixed regardless of the precise tilt value, only the intensity varies. The source photons will diffract off those 'crystallites' (perfect microcrystals within the mosaic crystal structure) which have the correct orientation and only experience normal absorption in the other crystallites. The intensity is maximal when $\tau = \theta_{Bragg}(E)$, and the image disappears completely when the deviation of $\tau$ from the ideal Bragg angle is much larger that the mosaic width, but the image never moves! The image position at energy, $E$, is completely specified by Bragg's law.

### 3) Lens platform and crystal configuration

We expect that the lens platform will be realized as a circular, ring-shaped aluminum honeycomb structure with an inner diameter of 1300 mm, an outer diameter of 3600 mm and a thickness of about 70 mm. The crystal adjustment monitoring capability, which is the central element in our concept, is realized through optical means. The availability of this system will allow us to relax the requirements for platform long-term stability significantly compared to the demands relevant for a classic Laue lens with fixed crystals. The relaxed requirements should result in a lower mass of the lens support panel, which at least partly should compensate for the added mass of the tilt pedestals.

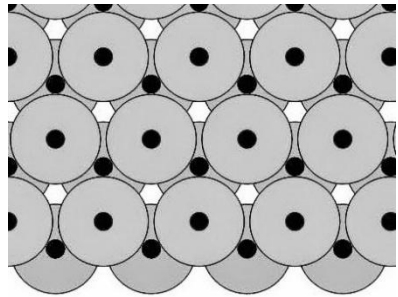

**Fig 3** *Double-layer crystal configuration. 65% of the area is covered by two overlapping crystal layers, 24% by one crystal layer. The small white hexagons are open (5% of the area), the black circular areas are obscured by the tilt-pedestals (6% of the area).*

The Laue crystals we plan to use are circular Silver and Copper disks, 40 mm diameter. Each crystal will be integrated in a 'crystal assembly' together with an optical alignment mirror. About 10,000 such assemblies will be employed. They are set in two layers, each layer arranged in a hexagonal pattern as illustrated in Figure 3 and 4. (The advantage of the double layer configuration is elaborated in paper I). The center-to-center separation of the crystals is 42.5 mm the size of the separation is dictated by the need to leave room for the pedestals on which the assemblies are mounted. The crystal thickness varies between 1.7 mm at the outer edge of the lens platform to 5.0 mm at the inner edge. The maximum mass of a single crystal is 104 g, the average mass is 32 g.

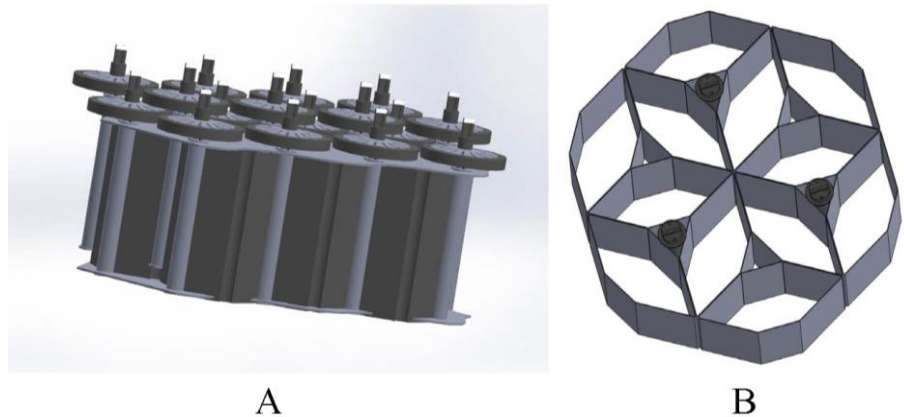

A                                      B

**Fig. 4** A: *Conceptual sketch of the planned lens panel structure with the tubular alignment pedestals embedded in the honeycomb panel. Two layers of crystals are shown. The crystal assemblies on top of the tilt pedestal carries the small alignment mirrors on the top. B: Concept study of the core structure of the honeycomb structure. The alignment pedestals shall occupy the triangular openings in the structure. Pedestals are shown in place for one layer of crystals.*

### 4) Tilt mechanism development

Our technology development program is divided into two main parts: A) development of a miniature tilt pedestal, and B) development of an optical alignment system.

As discussed above we require a tilt mechanism with one axis of adjustment only.

We summarize below our requirements for this mechanism:

Tilt range: +/- 1.5 degree. The tilt range actually needed in flight is only +/- 0.5 degrees, but we shall allow some margin for the initial crystal set-up. The Bragg angles for the lowest energies of interest, 150 keV, are just slightly over one degree. We expect the mechanical mounting of the crystal assemblies on the pedestals will ensure that the Bragg planes are aligned to better than 0.2 degrees relative to the pedestals. This is essentially a requirement on the crystal production methods: i.e. the surface of the finished crystals should be perpendicular to the Bragg plane to better than 10 arc-minutes. We also expect that we can mechanically align the optical reference mirror associated with each crystal to better than 5 arc-minute relative to the selected Bragg plane. (The finally achieved alignment must be known to better than a few arc-seconds)

Precision of tilt control: < 3 arc seconds.

Weight: less than 10 g – the weight of the mechanism should be significantly smaller than the average crystal mass.

Gamma-ray shadowing: The area covered by the pedestal should be less than 1 cm$^2$, the area must only be a small fraction of the 12.6 cm$^2$ area of each crystal.

Power: The mechanism must not consume any power except when actively adjusting the tilt angle.

Ruggedness: The mechanism should be capable of withstanding the typical satellite launch loads and vibration levels when supporting a crystal mass of 110 g.

On the basis of these constraints we have made a first design of a mechanism in the form of a slim cylindrical rod, with the crystal assembly riding on the top The cylindrical rod protrudes through the thick honeycomb lens panel, firmly guided on both sides. We aim for an 8 mm diameter rod corresponding to 4 % of the crystal area.

In order to minimize the dynamic forces on the tilt bearing, and the displacement amplitudes of the crystals caused by vibrations during the satellite launch, we demand that the center of gravity of the crystal assembly is placed exactly on the tilt axis. To do this we will need to cut an 8 mm central hole in each crystal. The crystal area we lose by making this hole is anyway not useful, since this area is shadowed by the tilt mechanism itself, but of course, the need for this hole is a complication for the crystal fabrication.

### 5) Mechanism design considerations

There are two critical elements in our tilt mechanism: the bearing and the 'motor'.

#### 5a) Tilt bearing

The first critical element is the tilt bearing. Conventional axle bearings are prone to permanently stick and lock into a fixed position. To avoid this problem we have made use of an early 20$^{\text{th}}$ century invention of a bearing relying on the flexure of crossed spring blades (Mesnager 1903). Such a flexible bearing is relatively easy to fabricate, has no slack and no stick, and is very rigid and strong against all displacement forces and bending moments, except around the axis defined by the crossed blades. We designed two miniature spring blades, which can be intertwined and glued into position to form the bearing, as illustrated in Figure 5.

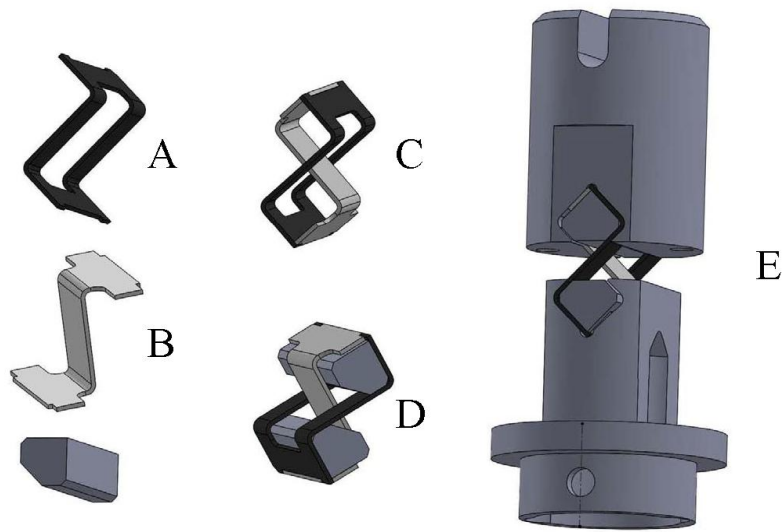

**Fig. 5** *Two crossed spring blades A and B, are the key elements of the flexible bearing. The steel blades are intertwined, C, and small filler pins are glued into the open spaces between the spring blades, D. At this stage the flexible bearing is completed. Finally, the flexible bearing is glued into the precisely shaped channels in the bearing socket (the fixed part) and the bearing head (the tilting part), E. The thickness of the steel blades is 0.15 mm and the width of the bearing proper is 5 mm. The separation between the socket and the bearing is 1.3 mm, so the free length of the steel blades is just under 2 mm.*

We have made load tests on four specimens of our bearings. We have subjected the bearings to loads of 20 N in the four orthogonal directions in the lens plane and to 40 N loads along the telescope axis. This corresponds to static accelerations of about 18 G transverse to the telescope axis and 36 G parallel to the axis even for the heaviest of our planned crystals. The bearing suffered no damage to the metallic parts during these load tests. However, in two of the four bearings the inserts (D in Figure 5) came loose when the 20 N loads were applied along the slotted channels in the bearing head. Obviously, we must improve on our assembly procedure to avoid this problem. Should this failure appear during launch, no mechanical parts will be released, but the flexible hinge may block and become inoperable.

Additionally, we have measured the flexure spring constants for our bearings in three orthogonal directions. If we take the measured spring constant against bending around the tilt axis as unity, then the spring constant for torsional twist around the lens axis is 35 times larger, and the spring constant for the third orthogonal direction is 60 times larger. Additionally, we have measured the natural frequency of a test specimen suspended on the bearing for oscillations around the tilt axis to be 7.1 Hz. The moment of inertia of the test specimen is 407 gcm$^2$. In table 1 we give the derived natural frequencies for crystals of different masses suspended on our flexible bearing.

**Table I. Natural frequencies for Laue crystals suspended on the flexible bearing**

|  | Crystal mass 30 g | Crystal mass 110 g |
|---|---|---|
| Free oscillations around the tilt axis | 26.2 Hz | 13.6 Hz |
| Torsional oscillations around the crystal axis | 110 Hz | 57 Hz |
| Oscillations around the 3$^{rd}$ axis | 203 Hz | 105 Hz |

Please note that in reality the oscillations around the tilt axis are not free, they will be heavily damped by the tilt mechanism described below.

The fact that there is no physical axle in the bearing means that there is no guarantee that the real turning axis is perfectly aligned with the line where the spring blades cross. Ideally, we demand that the axis of rotation is parallel to the lens plane and perpendicular to the radius vector from the telescope axis to the crystal center. Fortunately, due to the very small angular displacements we need, +/- 0.5 degrees after the initial set-up, the demands for the alignment of the ideal and the real axis of rotation are not very stringent. We have found that the most critical error is a tilt of the real axis relative to the lens plane where the tolerance is +/- 3 degrees. Misalignment of the real axis in the lens plane is less critical. This issue must be kept under control during the production of the bearings.

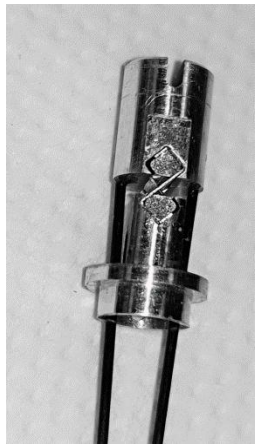

**Fig. 6** *Completed top part of the tilt mechanism with the crossed spring bearing in place. The upper cylindrical part is the moving part, which carries the crystal assembly, the lower cylinder with the flange is the fixed part connected to a 70 mm long tube, which protrudes into the honeycomb panel. The two black*

*carbon fiber rods (50 mm long) connect the top part rigidly to the tilt actuator. The weight of this part is 1.25 g*

### 5b) **Tilt actuator**

The second critical element in our tilt mechanism is the piezo-electric 'slip-stick' actuator, which performs the 'motor'-function and drives the tilt adjustments. A drawing of the actuator is shown in Figure 7. The 'slip-stick' mechanism relies on the difference between the static and the dynamic friction between a permanent magnet and an 'anchor' (Pohl 1987, Anders 1987, Hunstig 2017). The mechanism also relies on the inertia of the 0.3 g magnet in its brass holder. The anchor is set in vibration by the piezo-electric crystal. The vibrations are of asymmetric shape with either a sharp rise and a slow fall or the reverse. The magnet will stick to the anchor during the slow movements but slip during the fast movements. Repeated pulses will drive the magnet towards one end of the anchor. Reversing the shape of the electric pulses exciting the piezo will drive the magnet in the opposite sense. The magnet is rigidly connected via two 50 mm long carbon-fiber rods to the crystal bearing at the top of the tilt pedestal.

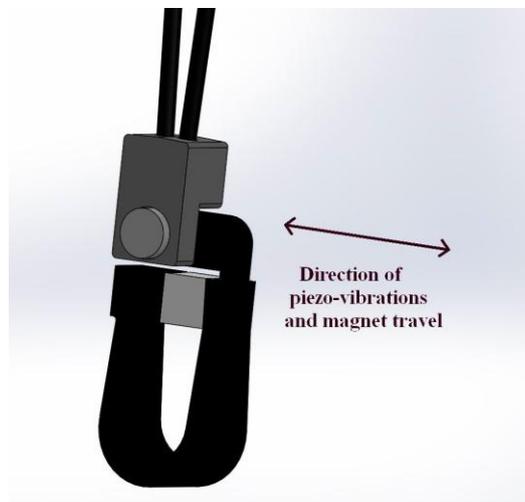

**Fig. 7** *The piezo electric actuator. The piezo crystal (light grey cube) is held in compression within a U-shaped steel spring (black). The piezo element needs to be under high compression (30 N) at all times for the mechanism to work. The cylindrical magnet slides back and forth on the black steel anchor flange at the top, dependent on asymmetrical, triangular drive pulses applied to the piezo crystal. The crystal is electrically isolated from the steel clamp by thin sheets of kapton film. Two 50 mm long carbon fibre rods connect to the tilting crystal bearing.*

The step size of the magnet movement for each triangular drive pulse is of order 0.001 mm, but may vary considerably from one pulse to the next, so an external reference system is required to arrive at a desired tilt angle. The actuator requires only one drive wire (and a current return path) and it does not require any power when not in action. The weight of our completed actuator is 1.55 g. We use commercially available, neodymium magnets of 2 mm diameter and 2 mm height with a holding force of 1 N, and piezo crystals from CTS corporation of dimensions 2x2x2 mm³. The piezo-crystal we use in our prototype requires a drive voltage of 60 V to provide the maximal displacement amplitude, but the mechanism works well with lower drive voltages, we have made successful tests down to 30 V drive pulses.

A cut through the complete design of the tilt mechanism including a 5 mm thick Laue crystal mounted on the top is shown in Figure 8. Note how the center of gravity of the Laue crystal is placed precisely at the level of the crossed spring tilt axis to minimize the forces on the bearing (and crystal oscillations) during launch.

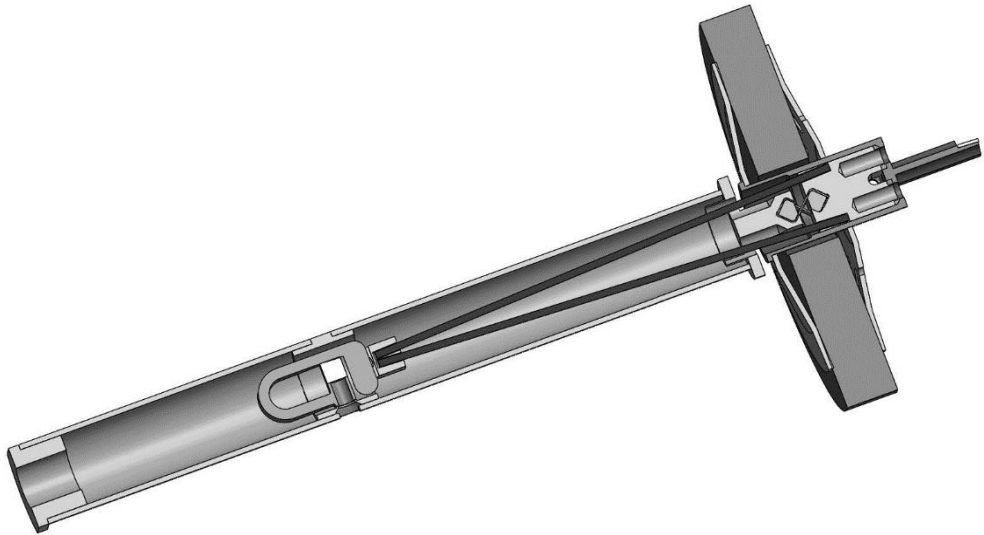

**Fig. 8** *Cut through the tilt mechanism design. with a thick Laue-crystal mounted on the top. A small alignment mirror (5x5 mm$^2$) is an integral part of the top crystal assembly. The detailed design of the top assembly is still open.*

### 6) Mechanism fabrication status

First versions of the two key elements of the tilt mechanism discussed above have been manufactured. A test set-up with both elements combined is shown in Figure 9.

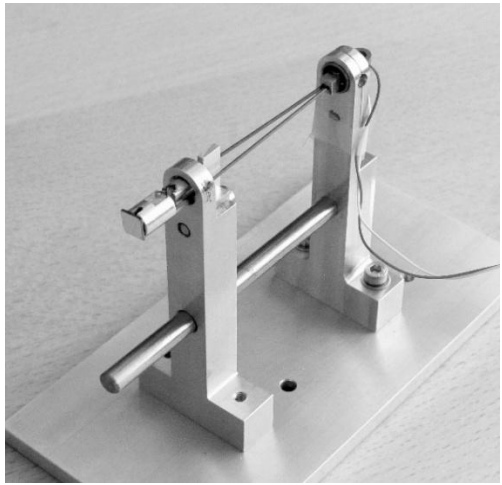

**Fig. 9** *Test set-up for the first versions of the piezo mechanism and the tilt bearing. The tilt bearing is in the front, the piezo drive at the back*

At this time we only have limited test results on our prototype. We have verified that the tilt range is ±1.5 degrees, as required. Moreover, we can adjust the tilt in sub-arc-second steps across the entire regulation range. This is where the no-stick flexible bearing is really showing its worth: even the most minute piezo movements are immediately reflected in the crystal tilt. The mechanism can move across the entire angular range in about one second. The resolution of the mechanism is better than one arc second, so it will be the resolution of the optical alignment system which will define the precision of the Bragg angle adjustments.

We still need to verify our planned design for the crystal assembly, which will support the very fragile Laue-crystals through the launch ordeal. On top of this assembly is the small reference mirror attached to each crystal. It is the angle between this mirror normal and the Bragg plane normal, which must be determined with arc-second precision during the integration of these crystal assemblies. The Laue-crystals are extremely soft, so we plan to keep the crystals in place using purely mechanical contact between a rigid under-plate and a flexible over-plate holding the crystal in place. We are aware that the local contact pressure must be small (less than 10 N/cm$^2$) and well controlled.

All in all, we are still optimistic that we can complete the tilt platform within our 10 g weight budget.

### 7) Optical alignment system

This system is still only in the design phase, however, a functioning proof-of-principle version of the system has already been set up in our laboratory.

As indicated in Figure 1 we intend to make use of a set of three auto-collimator telescopes placed on a turret in the center of the Laue lens. In this description 'auto-collimator' just denotes an optical system in which a point light source is located in the focal plane of the telescope close to the optical axis. The divergent beam from the light source inside the telescope emerges as a parallel light beam after passing through the telescope objective. The auto-collimator system is inclined slightly downward with respect to the lens platform, and the beam will illuminate a small number of crystals on the platform, as shown in Figure 10A.

The small (5x5 mm$^2$) alignment mirrors associated with each crystal will reflect back the light towards the auto-collimator. We will use autocollimator telescopes with large apertures (150 mm diameter). With this aperture between 20 and 35 alignment mirrors may simultaneously send back their light beam toward the autocollimator front lens. (Some of the returning beams may miss the lens – the corresponding crystals must be aligned from a different autocollimator azimuthal orientation). Each mirror beam returning to the autocollimator will be focused to a small spot on a CCD image sensor in the focal plane.

It is the vertical position of the resulting light spots on the CCD, which will be used to check the tilt angle of the alignment mirrors – and thereby the Bragg-angle of the Laue crystals. In the horizontal (azimuthal) direction the position of the spots will depend on the instantaneous orientation of the autocollimator platform, but the relative ordering of the spots in this direction is constant, independent of any autocollimator rotations or crystal tilt changes.

We will need to assure in the design that the light spots in the focal plane do not overlap each other since this could affect the precision with which we can locate the centroid of each spot. We plan to assure the separation of the light spots by assigning each of the 10000 crystals in the lens a specific slice in azimuth (each slice 2.16 arc-minutes wide) and establish an assembly procedure, which assures the fulfillment of this constraint. Some preliminary considerations on this procedure are explained in the Appendix.

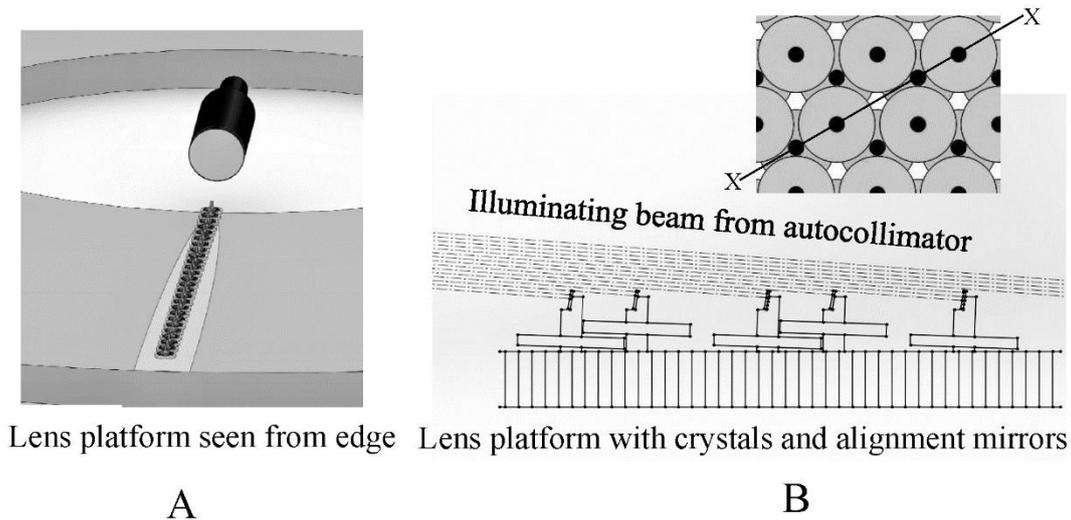

Lens platform seen from edge    Lens platform with crystals and alignment mirrors

A                                          B

**Fig. 10** A: *The beam of parallel light emerging from the auto-collimator telescope illuminates a narrow stripe of crystals on the lens platform. B: The beam is tilted down relative to the lens panel. The downward inclination will ensure that the alignment mirrors close to the telescope axis do not completely block the illumination of the mirrors located further out. The track X-X delineates the most critical direction for the mirror-mirror shado*wing.

The alignment mirrors close to the Laue-lens axis must not shadow other mirrors further out. Therefore, the optical axis of the auto-collimator is tilted down by 5º relative to the plane of the lens panel, as indicated in Figure 10B. This is just enough to limit the possible shadowing to an acceptable degree, even for the mirrors with the smallest separation in the double layer lens. (At least a mirror area of 2x5 mm$^2$ will remain un-shadowed). The mirrors close to the center of the platform must also not shadow the returning light beams coming from the outer mirror on the lens. Thus, also the alignment mirrors must be tilted up by 5º relative to the Bragg plane of their associated crystal.

The auto-collimator telescope rotates in small steps around the telescope axis and scan the entire lens surface to verify the correct alignment of all 10,000 crystals. It is the rotation axis of the auto-collimator system, which defines the gamma ray axis of the Laue lens, and it *w*ill be the stability of this system, which determines the ultimate accuracy of the crystal tilt alignment.

We emphasize that the use of the optical tilt adjustment technique will significantly relax the requirements on the rigidity and stability of the honeycomb platform supporting the Laue lens. The angular alignment of the crystals is independent of small displacements or distortions of the mounting platform. The relaxation of the stiffness and long-term stability requirements will certainly influence the mass and the cost of this item. Short term deformations caused by temperature variations must still be kept well under control.

The alignment process will have to be repeated every time we need to re-tune the Laue lens for a new energy range. Because the alignment system is indispensable for the mission success, redundancy must be assured in the design. We plan to have three auto-collimator telescopes in hot redundancy and working in parallel during the retuning process. We expect that the time spent to retune one crystal is about 1 s. This would imply that a complete re-tuning can be completed in under an hour using all three autocollimators.

## 8) Summary

We have designed and fabricated the key elements of a miniature alignment pedestal intended for use with individual crystals in a Laue lens. The initial test results are very encouraging: the tilt angle can be controlled in arc-second steps, and the mechanism can sweep through its full angular range in less than one second.

The pedestal is 8 mm in diameter and will therefore only block the central 4 % of a 40 mm diameter Laue crystal. We intend to cut an 8 mm hole at the center of each crystal to allow for a robust mechanical mounting of the crystal assemblies on the alignment pedestals.

Our original idea was to use the alignment pedestals just to facilitate the final alignment of the 10,000 crystals on the lens. However, we have realized that the availability of the alignment pedestals will allow the re-tuning of the complete lens for a new energy range. Simulations have shown that this opens new possibilities to use a single Laue lens to study astrophysical sources over a much wider energy range than hitherto thought possible (see paper I).

To achieve the required arc-second alignment accuracy an external reference system is required. We plan to realize this using a system of optical auto-collimators mounted on a scanning platform in the center of the Laue lens. In this way, the entire lens will be aligned relative to the rotation axis of the auto-collimator platform. This construction will considerably relax the requirements for the long-term mechanical stability of the support platform for the Laue crystals.

The special crystal mount provided by the pedestals also opens the possibility to stack two overlapping crystal layers on the same lens platform. This increases the lens collection power by about 65 % (paper I).

## 9) Conclusions

A fully tunable Laue lens now appears to be a realistic proposition. Such a lens will provide exciting new possibilities to attack the nuclear line energy range between 200 keV and 2.2 MeV. The simulations on a specific lens design compatible with current standard launch vehicles indicate that continuum sensitivities better than $10^{-7}$ photons/cm$^2$keV can be obtained with observation times of a few days.

## 10) Acknowledgements

I am indebted to DTU Space for supporting this development work. I am indebted to Josef Polny and Niels Christian Jessen for support with the finite element calculations and the design of the piezo actuator. I am also indebted to the staff of the DTU central mechanical workshop for useful discussions and for their dedicated and skilled work with the minute mechanical parts required for this project.

## 11) References

Anders, M. et al., (1987), Surface Science, 181, p 176

Hunstig, M. (2017), Actuator, 6, p 7

Mesnager, A., (1903), Sur les articulations à lames flexibles. Comptes Rendus Hebdomadaires des Séances de l'Académie des Sciences, 137, p. 908

Lund, N., (2020), Wide band, tunable gamma-ray lenses (arXiv xxxxxxx)

Pohl, D., (1987), Rev. Sci. Instr., 58, p 54

## Appendix

### Azimuthal orientation of the alignment mirrors

It is important for the accuracy of the crystal alignment that the individual mirror images in the focal plane of the autocollimator telescopes are kept well separated. In principle we could rely on the separation both in tilt and azimuth. However, since the azimuth of the mirrors is fixed whereas the tilt is adjustable it is appealing to try to separate the mirrors in azimuth during the lens assembly. If this can be arranged, it is done once and for all, and we don't need to worry about the image separation anymore.

Our nominal concept was to orient all the mirrors antiparallel to the radius vector from the Laue lens center to the center of each crystal. However, this would imply that a significant fraction of the mirrors would be too close to their neighbors in the azimuth angle, and a few hundred would even line up completely.

We have about 10000 mirrors to care for so we can allocate an azimuth interval of 2.1 arc minutes to each. With a image size in the focal plane of less than 30 arc-seconds this is adequate to reliably separate and identify individual image spots.

For every crystal in the lens it is straightforward to determine the nominal azimuthal orientation of the mirror (antiparallel to the relevant radius vector) and therefore also to determine the shift required to achieve a uniform separation in azimuth of 2.1 arc-minutes between all neighboring mirrors.

We assume that during the integration of the crystal assemblies we can fix the alignment mirrors with a mean error of about 2 arc-minute relative to the Bragg plane of the Laue crystal. We will afterwards measure the relative offset of these two surfaces to better than a few arc-seconds. With these data available for all crystal assemblies we can for each position on the lens surface select that asembly, which best matches the actual and the desired offset. We estimate that this procedure will suffice for almost 90% of the crystal assemblies. Recall, that as discussed in section 2 we do have a bit of freedom (a few arc-minutes) in the alignment of the Bragg plane azimuth.

For the remaining about 10 % of the assemblies we must use a specific assembly integration procedure to ensure that when the crystal is installed on the Laue platform and turned so the mirror normal is placed inside its desired azimuth interval, then the Bragg plane is correctly oriented for the diffracted image to stay inside the gamma-detector. This may sound complicated, but the lens assembly will certainly be a fully computer controlled activity, so every crystal assembly will be assigned a unique identification number, and thus a unique position on the lens platform. It will therefore be manageable to ensure that this azimuth constraint is fulfilled.